# Yield-stress anomaly in equiatomic ZrNbTiVHf high-entropy alloys

T. Lienig, C. Thomas, M. Feuerbacher
*Ernst Ruska-Centre for Microscopy and Spectroscopy with Electrons and Peter Grünberg Institute, Forschungszentrum Jülich GmbH, 52425 Jülich, Germany*

**Abstract**
We have carried out plastic deformation experiments on single-phase samples of the body centered ZrNbTiVHf high-entropy alloy between room temperature and 1150 K. The experiments were carried out on polycrystalline samples with two different grain sizes, in compression at a true constant strain rate of $10^{-4}$ 1/s. Incremental tests such as stress-relaxation tests, strain-rate changes and temperature changes were carried out in order to determine thermodynamic activation parameters of the plastic deformation process. The material displays a yield-stress anomaly in the temperature range between about 500 and 800 K, where the yield stress increases with increasing temperature. In the same temperature range, we find an extremely low strain-rate dependence of the flow stress, which is reflected in almost constant flow-stress values in stress-relaxations and strain-rate changes. At temperatures below 400 and above 800 K we find regular plastic deformation behavior with a linear temperature dependence of the yield stress of about -1.1 MPa/K, an activation volume of around 1.8 nm$^3$ and 0.2 nm$^3$ at low and high stresses, respectively, and an activation enthalpy of 2.7 eV in the high-temperature range.

**Introduction**
High-entropy alloys (HEAs) are materials composed of multiple metallic elements, typically 5, in equiatomic or near-equiatomic composition so that none of the elements acts as a dominating alloy basis [1, 2]. Ideally, HEAs solidify as a random solid solution on a simple bcc, fcc or hcp [3] crystal lattice. Therewith they possess a unique combination of chemical disorder and topological order and establish a novel type of solid matter in between metallic glasses and ordered crystals.
Numerous studies have been devoted to the mechanical properties of HEAs, but the consequences of their salient structural features for the plastic behavior to date are far from being understood. The determination of the intrinsic deformation properties of HEAs requires dedicated experiments on well-characterized, single phase samples. In the present paper, we report on a deformation study of the equiatomic ZrNbTiVHf HEA, a homogeneous single-phased solid solution with a bcc crystal lattice. We have measured stress-strain curves in a wide temperature range from room temperature to 1150 K and determined the relevant thermodynamic activation parameters of the plastic deformation mechanism. We compare our results with those on other, related HEAs and on known intermetallic phases.

**Experimental**
High purity elements Zr, Nb, Ti, V and Hf in equiatomic proportion were arc molten and cast into a copper mold. In order to achieve a high homogeneity, the ingots were remolten several times in a levitation furnace and eventually quenched into a water-cooled cylindrical copper mold of 10 mm in diameter. Subsequently, the rods were annealed under argon at 927 K for 2 h and 48 h.

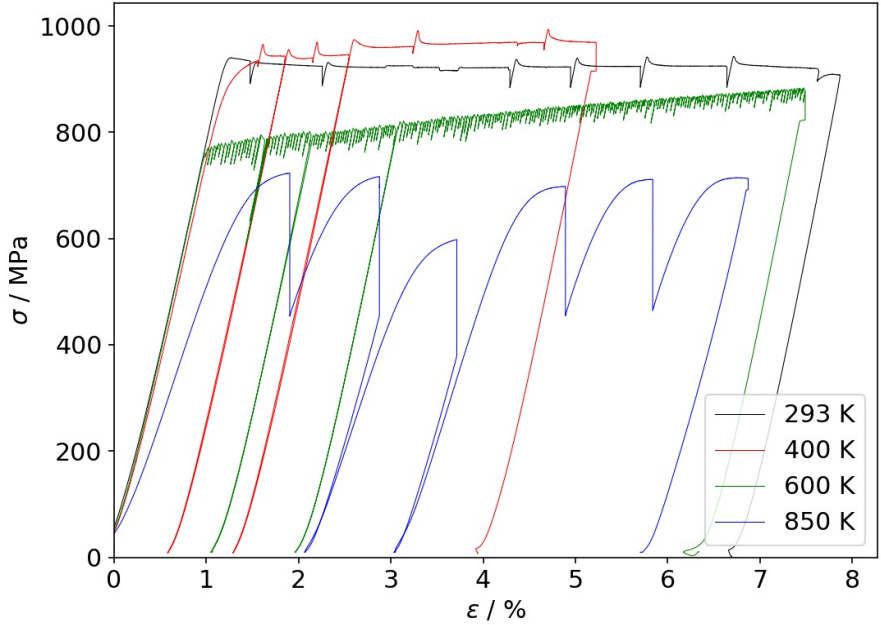

*Fig. 1: True-stress vs. true-strain curves at four different temperatures of the coarse grained ZrNbTiVHf HEA.*

The resulting material is a polycrystalline, single phase solid-solution based on a bcc crystal lattice. Details of the materials preparation and characterization are provided elsewhere [4]. The average diameter of the single grains was determined as about 1 mm and 300 to 400 µm for the material annealed for 48 h and 2 h, respectively. In the following we will refer to these materials as coarse grained and finer grained, respectively.

For the plastic deformation experiments samples of rectangular cuboidal shape of about 1.8 x 1.8 x 4 mm$^3$ were cut by spark erosion and subsequently ground using 1200 grit SiC paper. Samples were prepared from both the coarse and the finer grained material. The respective single grain sizes translate to an average amount of grains per deformation sample of the order of 10 for the coarse grained and up to about 500 for the finer grained material.

Plastic deformation experiments were performed in compression along the long sample axis in a modified ZWICK Z050 uniaxial testing system at constant true strain rate of $10^{-4}$ s$^{-1}$ under closed-loop control at temperatures between room temperature and 1150 K. The strain was measured directly at the sample by a linear inductive differential transducer at an accuracy of ±10 nm. Our deformation experiments included temperature-cycling tests, stress-relaxation tests and strain-rate changes in order to determine thermodynamic activation parameters of the deformation process. As a basic testing scheme, we deform the sample at a given temperature slightly further than the yield point, carry out a stress-relaxation test for 120 seconds and subsequently unload the sample for a temperature change. The temperature is raised by 20 K with the sample held at a constant small load of 10 to 20 N until thermal equilibrium is reached, which takes about 20 to 30 minutes. The sample is then reloaded and a second stress-relaxation and subsequent temperature change to the initial temperature is carried out. After a second reload we carry out another stress-relaxation and deform the sample up to a total strain of about 6 to 8 %. Typically we carry out additional stress relaxations and strain-rate changes (up and down by 20 %, i.e. to 1.2 and 0.8 x $10^{-4}$ s$^{-1}$) along this course, to obtain more data and check the reproducibility and consistency. This allows us to determine the activation volume, stress exponent and activation enthalpy at different values of stress, strain and temperature according to the evaluation procedures provided e.g. in [5].

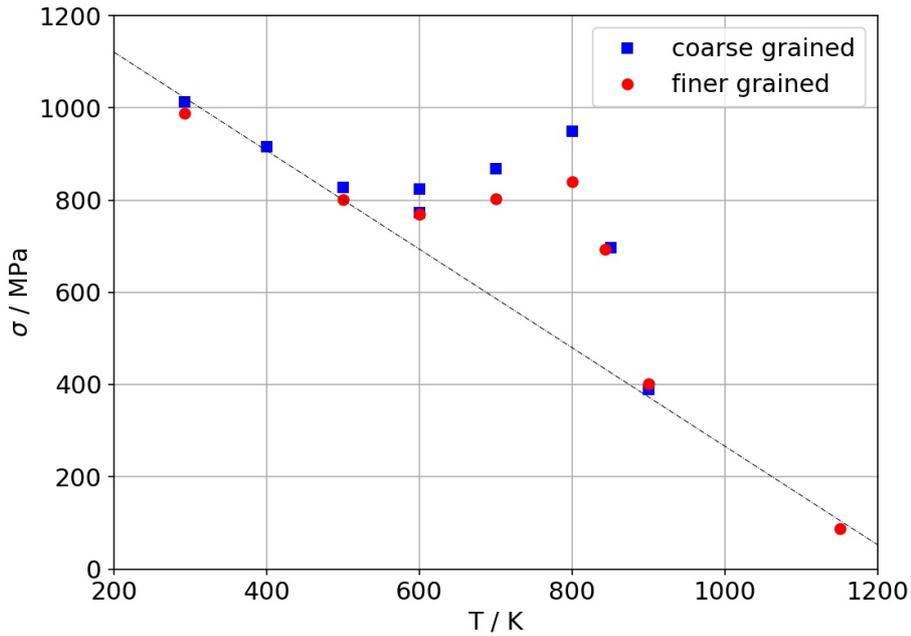

*Fig. 2: The 0.2 % yield stress as a function of temperature T for the coarse grained (blue squares) and finer grained (red circles) ZrNbTiVHf HEA.*

**Results**

Fig. 1 shows representative true-stress true-strain curves $\sigma(\varepsilon)$ of the coarse grained samples at four different temperatures. All curves follow the basic testing scheme except for the curve at room temperature, at which no temperature changes were carried out. All curves show an initial elastic loading stage, leading into plastic yielding without displaying a pronounced yield point. Stress-relaxation tests are visible as vertical dips in the curves, and strongly vary in depth for the different temperatures. Temperature changes are preceded and followed by unloading and reloading segments and lead to stress changes, which for the shown curves only at the highest temperature are of appreciable magnitude. At the lower temperatures, reloading segments are followed by distinct stress overshoots. It is furthermore noticeable that the medium temperature curve at 600 K displays stress fluctuations at fairly regular strain intervals and depth. Such behavior is found for the deformations at 500, 600 and 700 K. At 600 K stress drops of about 20 MPa occur parallel to the elastic loading segment at strain intervals of 0.03 %. The stress-strain curves of the finer grained sample are qualitatively and quantitatively very similar.

Fig. 2 displays the 0.2 % yield stress $\sigma_{0.2}$ as a function of temperature $T$ for the coarse grained (blue squares) and finer grained (red circles) samples. At the lowest and highest temperatures the yield stress is decreasing with increasing temperature, which reflects the normal behavior observed for most metallic materials. The dashed line in Fig. 2 is a guide to the eye following a rate of 1.1 MPa / K. In the temperature range between about 500 to 800 K, however, the yield stress increases with increasing temperature, which is considered anomalous yielding behavior. The yield stress and its temperature dependence of the coarse and finer grained material are similar, but at lower temperatures the coarse grained material displays a higher yield stress, the difference increasing with increasing temperature and peaking at about 10 % at 800 K.

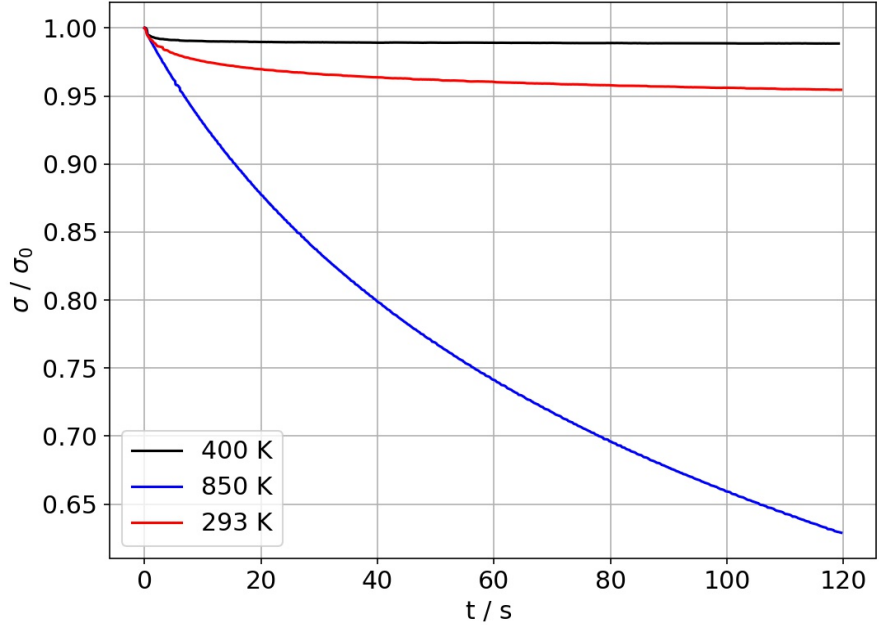

*Fig. 3: Stress-relaxation curves normalized to the initial stress value of the ZrNbTiVHf HEA at three different temperatures.*

The activation volume $V$ of the plastic deformation process was mainly determined by stress-relaxation tests. In the course of these tests, the total strain is held constant for an amount of time, 120 seconds in our case, during which the variation of the stress over time is measured. Three stress-relaxation curves obtained at three different temperatures are displayed in Fig. 3. The stress data is scaled by the initial value at the start of the relaxation and is plotted as a function of the time $t$ into the relaxation test. The curve at 293 K is typical for the low-temperature range and displays a moderate decrease of the stress of about 5 % over the total relaxation time and the typical logarithmic [6] time dependence of a metallic material. In the high-temperature range, represented by the 850 K curve, a qualitatively similar time dependence, albeit with a much stronger stress decrease totaling at over 35 % after 120 seconds is found. The relaxation in the intermediate temperature range, corresponding to the anomalous yielding range in Fig. 2, shows a completely different behavior. Here the stress, after an initial short drop at the beginning of the relaxation, is essentially constant. This behavior is also reflected directly in the stress-strain curves (Fig.1), where the vertical drop due to the stress relaxations is clearly discernible at room temperature and very distinct at the high temperature, but hardly visible in the intermediate temperature range.

The stress-relaxation data is further evaluated following the procedure described e.g. in [5], by determining the slope of the data plotted as the negative time derivative of the stress versus the stress. From the slope, the activation volume can be directly calculated, assuming a Schmid factor of 0.5. For comparison with the literature, we additionally provide the corresponding values of the strain-rate sensitivity SRS [7], which directly equals the inverse of the slope of the $\ln(-\dot{\sigma})$ vs. $\sigma$ plot, and hence requires no further assumptions for calculation. Our experimentally obtained values of the activation volume are summarized in Fig. 4. Red symbols represent values in the high-temperature range, i.e. from experiments at 800 K and higher. The absolute values obtained are about 1.8 nm$^3$ at low stress and about 0.1 to 0.2 nm$^3$ at higher stress. This corresponds to SRS values of 15 MPa at low stress, and 200 to 100 MPa in the high-stress region, respectively. The stress dependence of the activation volume follows a hyperbolic behavior (dash-dotted line). We see no systematic difference between the values obtained from the coarse and finer grained material, the results are identical within the scatter of the experiment. The values in the low-temperature range are distinctly different, we find an activation volume of about 1 nm$^3$, which corresponds to a strain-rate sensitivity of about 8 MPa. Since the low-temperature range accessible to our experimental equipment is rather narrow, the stress range is also very limited. We can cover values from about 900 to 1100 MPa, which display an essentially constant activation volume, with a slight tendency to decrease with increasing stress. In this temperature range, again we see no significant difference between coarse and finer grained material.

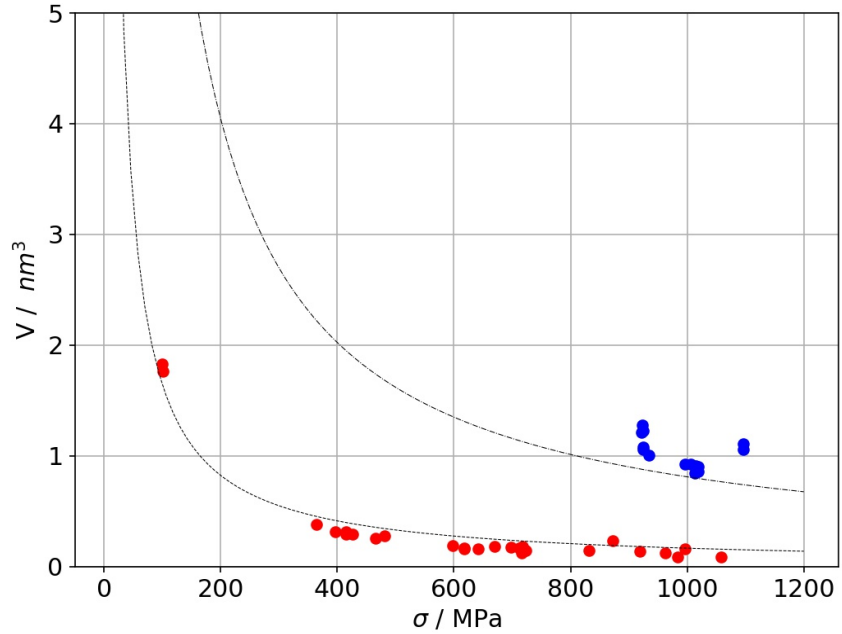

*Fig. 4: Stress dependence of the activation volume of the ZrNbTiVHf HEA in the low-temperature (blue) and high-temperature range (red). The dash-dotted and dashed lines are fits of a hyperbolical function to the low- and high-temperature data, respectively.*

In the intermediate temperature range, the stress-relaxation data cannot be evaluated following the same procedure, since the stress drop is extremely small and does not follow a logarithmic behavior. We can only assert that the strain-rate sensitivity of the flow stress is very small, which corresponds to the observation that the strain-rate changes conducted in the intermediate range do not lead to a discernible change of the flow stress.

The stress exponent was determined from the same stress-relaxation data, fitting the slope in a $\ln(-\dot{\sigma})$ vs. $\ln(\sigma)$ representation [5]. In the high-temperature range we obtain values between about 4.0 and 4.7, averaging to 4.4 with no significant difference between coarse and finer grained material. In the low-temperature range, on the other hand, we obtain very large values between 90 and 110.

The activation enthalpy of the deformation mechanism could be determined in the high-temperature range from the simultaneous evaluation of stress-relaxations and temperature changes [5]. We obtain a value of 2.7 ± 0.5 eV averaged over the experiments with coarse and finer grained samples in the temperature range of 800 to 900 K, with no significant temperature dependence. In the intermediate temperature range no value for the activation enthalpy could be determined, since no values for the activation volume could be evaluated and the temperature changes did not lead to appreciable changes of the flow stress. In consistency with the observed yield-stress anomaly in the intermediate temperature range, we observed a small but positive stress change, e.g. of 0.4 MPa/K at 700 K.

**Discussion**

In this paper we report on the macroscopic plastic deformation properties of polycrystalline ZrNbTiVHf HEAs in the temperature range from 293 to 1150 K. With a melting temperature of 1863 K [4], the high-temperature end of our experiments corresponds to a homologous temperature of 0.62. Our experiments were carried out in compression up to strains of 6 to 8 %, with a focus on thermodynamic activation analysis, and without the intention to determine the ductility limits of the material.

At room temperature we find a yield stress of about 1 GPa and at 1150 K of about 90 MPa. The most striking feature of the plasticity of the ZrNbTiVHf HEA alloy is the presence of a yield-stress anomaly (YSA) in the range from about 500 to 800 K. Here, the yield stress of the material increases with increasing deformation temperature, which is considered anomalous behavior for a metallic material. Outside this temperature range and to the extent of the range investigated, the material behaves normally, i.e. the stress decreases with increasing deformation temperature. The presence of a YSA has been observed for several ordered intermetallic alloys such as Ni-Al , Cu-Al, Cu-Zn and Ti-Co, but also in disordered alloys and pure metals [8]. This paper, for the first time, reports on the observation of a YSA in a high-entropy alloy.

In the temperature range 500 to 700 K the stress-strain curves display strong variations in the flow stress, which is in distinct contrast to the very smooth curves in the low- and high-temperature range. This indicates the presence of a flow-stress instability and resulting serrations of the flow stress in that temperature range. We can, however, not exclude that the observed flow-stress variations are due to noisy behavior of the feedback-loop controlling the flow stress under these conditions of stress and temperature. Further experiments to clarify this issue are currently underway. However, currently we tend to believe that the measured flow-stress variations are real serrations, due to the following arguments: i) the stress increments are very consistent and appear after regular strain increments, ii) the downward fluctuations of the flow stress are parallel to the elastic unloading curves, iii) the stress variations occur in the exact same temperature range, where we also find an extremely low strain-rate dependence of the flow stress. Serrated stress-strain curves have been previously observed in FeCoCrMnNi HEAs [9] at 673 and 873 K.

We find that the deformation behavior of the material is distinctly different in the low-, high- and intermediate temperature range of our experiments. Besides the yield stress, this is reflected in the response of the material to stress-relaxation, strain-rate changes and temperature changes, and in the smoothness of the flow stress. On the basis of these observations, we distinguish three different temperature ranges in the deformation behavior of the ZrNbTiVHf high-entropy alloy with the following characteristics:

*Low-temperature range, room temperature to about 400 K*
- The yield stress decreases with increasing temperature
- Very little or no grain-size dependence of the yield stress
- Stress overshoots after reloading segments
- In stress-relaxation experiments, the stress decreases with time in a "regular" logarithmic way

- Activation volume of about 1 nm$^3$
- Stress exponent of the order of 100
- Upon strain-rate changes, the material shows appreciable stress response, the amount of which corresponds to the activation volume determined by stress relaxation
- Smooth flow stress

*Intermediate temperature range, about 400 to 800 K:*
- Presence of a YSA: the yield stress is constant or increasing with increasing temperature
- The yield stress of the coarser grained samples is higher up to about 10 %
- No or very little response of the flow stress to strain-rate or temperature changes
- After a small initial drop, the stress remains constant in stress-relaxation tests
- The intermediate temperature range contains a smaller range from 500 K to 700 K where the flow stress is apparently serrated

*High-temperature range above about 800 K*
- The yield stress decreases with increasing temperature
- Very little or no grain-size dependence of the yield stress
- In stress-relaxation experiments, the stress decreases with time in a "regular" logarithmic way, and much stronger than in the low-temperature range
- The activation volume shows a hyperbolic stress dependence and amounts to about 1.8 nm$^3$ at 100 MPa and about 0.1 nm$^3$ at 1 GPa
- Stress exponent of about 4.4
- Strong response of the flow stress to strain-rate and temperature changes
- The activation enthalpy could be determined and averages to about 2.7 eV
- Smooth flow stress

The high-temperature range, due to the strong temperature dependence of the flow stress in that range, includes a wide stress range, which allowed us to determine the stress dependence of the activation volume. The data can be fitted to a hyperbolic function (dashed line in Fig. 4) using a single parameter, with a best fit at 164 nm$^3$/MPa. Due to the relationship between the stress exponent and the activation volume [5], the fit parameter can be translated to a stress exponent of 5.95. This fits reasonably well with a direct determination of the stress exponent form stress-relaxation test, which in the high-temperature range averages to 4.4. Likewise, a hyperbolic function calculated using a stress exponent of 100 (dashed-dotted line in Fig. 4) is fairly consistent with our low-temperature relaxation data. This shows that our data and our analysis are consistent.

The stress exponent of 4.4 corresponds to frequently observed values in metallic materials. On the other hand, the low-temperature value of the order of 100 is very large and suggests that the description of the deformation behavior in the form of a simple power law is not adequate in this temperature range.

At low temperatures, due to the small temperature range and the rather small temperature dependence of the flow stress, we cannot evaluate the stress dependence of the activation volume. Extension of the exprimental range to lower temperature is surely possible with adequate equipment, but will prospectively not increase the range to lower stresses.

The only previously published study of the plasticity of equiatomic ZrNbTiVHf HEAs was carried out by Fazakas et al. [10]. These authors deformed polycrystalline samples in compression at a strain rate of 5 x 10$^{-4}$ s$^{-1}$ at room temperature. They find a high ductility (up to 40 % for material heat treated at 1173 K) and a yield stress of about 1400 MPa. The yield stress is thus considerably higher than that observed in the present study, which may be due to the higher strain rate in that study. The results of very recent tensile-testing tests on as-grown ZrNbTiVHf by Gadelmayer et al. [11] are in good agreement with those of the present study, and confirm the presence of a YSA in the temperature range of 500 to 800 K.

A number of studies on the plastic behavior of other refractory HEAs have been published to date. Senkov et al [12] investigated the deformation behavior of equiatomic ZrNbTiTaHf polycrystals. As the HEA investigated in the present study, ZrNbTiTaHf is a single phase bcc solid solution and most likely the two materials are related by direct substitution of Ta and V. Nevertheless their plastic properties significantly differ in that the yield stress of ZrNbTiTaHf strictly monotonically decreases over the entire temperature range from 293 to 1473 K, and there is no indication of temperature induced changes in deformation mechanism. Similar results were obtained by Feuerbacher et al. [13] in the smaller temperature range from 293 to 573 K. Feuerbacher et al. [13] also report the observation of low-temperature stress overshoots in ZrNbTiTaHf after reloading, as found for ZrNbTiVHf in the present study.

Couzinié et al. [14] studied the room temperature plasticity of ZrNbTiTaHf HEAs and carried out stress-relaxation tests to determine the activation volume. They find a yield stress of 905 MPa and an activation volume of 42 to 48 $b^3$ at strains between 1 and 3 %, where the activation volume was scaled by the cube of the b = a/2 <111> Burgers vector. Scaling the activation volume found for room temperature deformation for the ZrNbTiVHf HEA in the same way using a lattice parameter of 0.336 nm [13], we obtain values in the range of 40 to 50 $b^3$. Thus, the results of Couzinié et al. [14] correspond very well to the room-temperature results of the present study.

Senkov et al [12] also investigated two high-melting HEAs, TaNbMoW and TaNbVMoW. The temperature dependence of the yield stress of these materials does not decrease strictly monotonically with temperature, and displays the presence of a slight plateau between 800 and 1300 K. This may indicate the presence of a YSA in these materials as well. Whether or not this is the case, may be figured out by further deformation experiments at a lower strain rate.

The origin of the YSA in the ZrNbTiVHf HEA is currently unclear. We are presently undertaking further deformation experiments under different conditions and are carrying out microstructural and dislocation analyses of deformed samples in the transmission electron microscope. For the case of conventional intermetallic alloys, different mechanisms leading to the presence of a YSA have been identified, such as Kear-Wilsdorf locks and antiphase-boundary jumps in ordered phases, as well as dislocation locking mechanisms controlled by a diffusion mechanism [8]. The latter mechanism typically involves flow-stress instabilities, which for the ZrNbTiVHf HEA also appear to be present in the temperature range of the anomaly. Furthermore, it is known that phase transitions can also lead to YSA effects, which should be taken into account for the present material as well, as thermodynamic calculations using a CALPHAD approach indicated the presence of competing low-temperature phases in the Zr-Nb-Ti-V-Hf system [15].

**Acknowledgments**
The authors thank C. Gadelmayer, S. Haas, and U. Glatzel (U. Bayreuth) for inspiring discussions, and D. Kaiser and J. Heinen (FZ Jülich) for assistance with the data analysis. This work was financially supported by the German Science Foundation (DFG) under Grant No. FE 571/4-1.